# Superconducting properties variation with A15 composition in Nb$_3$Sn

Yingxu Li [a)] and Yuanwen Gao

*Key Laboratory of Mechanics on Environment and Disaster in Western China, The Ministry of Education of China, Lanzhou, Gansu 730000, P. R. China*

*Department of Mechanics and Engineering Science, College of Civil Engineering and Mechanics, Lanzhou University, Lanzhou, Gansu 730000, P. R. China*

[a)] *Corresponding author: liyx2010@lzu.edu.cn; Tel: 0086 931 891 4359; Fax: 0086 931 891 4561.*

PACS: 74.70.Ad; 74.25.fc; 74.20.Fg.

**Abstract**: We extend the Ginzburg-Landau-Abrikosov-Gor'kov (GLAG) theory to account for the variation of the upper critical field $H_{c2}$ with Sn content in A15-type Nb$_3$Sn. $H_{c2}$ at the vicinity of the critical temperature $T_c$ is related quantitatively to the electrical resistivity, specific heat capacity coefficient and $T_c$, with inclusion of the electron-phonon coupling correction, Pauli paramagnetic limiting and martensitic phase transformation of A15 lattices. $H_{c2}$ near $T_c$ is then extrapolated to $H_{c2}(0)$ at temperature 0K, and $H_{c2}(0)$ versus tin content agrees well with experiment results. We find that, as Sn content gradually approaches the stoichiometry, Nb$_3$Sn undergoes a transition from the dirty limit to clean limit, split by the phase transformation boundary. The *H-T* phase boundary and the pinning force show different behaviors in the cubic and tetragonal phase. Applying the theoretical formulas in technical Nb$_3$Sn wires, we obtain the dependence of the composition gradient on the superconducting properties variation in the A15 layer, as well as the curved tail at vicinity of $H_{c2}$ in the Kramer plot of the Nb$_3$Sn wire. This gives a better understanding of the inhomogeneous-composition inducing discrepancy between the results by the state-of-art scaling laws and experiments.

**Keywords:** Upper critical field; Composition inhomogeneity; Superconductivity; A15 type Nb$_3$Sn.

## 1. Introduction

At present, conventional low-temperature superconductors such as Nb$_3$Sn have been extensively applied in high-energy and nuclear physics, as well as in magnetic resonance imaging systems [1]. Inhomogeneity of Sn content is inevitable in practical Nb$_3$Sn conductors, due to the high vapor pressure of Sn at the formation temperature of the A15 phase in a solid-state diffusion reaction [2, 3]. The tin variation in a conductor covers nearly the entire A15 phase field of binary Nb$_{1-\beta}$Sn$_\beta$ with $\beta = 0.18 \sim 0.255$ [4]. In a Nb$_3$Sn conductor, the Sn gradient across the A15 layer has a remarkable impact on the local variation of the superconducting properties. Experiments show that, the upper critical field $B_{c2}$ varies almost linearly at ~5T per at% between 19.5% and ~24 at% Sn, and the transition temperature $T_c$ versus tin concentration $\beta$ also exhibits a linear relation within nearly the entire A15 phase field [3]. When $\beta$ approaching stoichiometry of Nb$_3$Sn, $\beta \approx 24$at%, $T_c$ (or $B_{c2}$) versus $\beta$ no longer keeps the linear relationship due to the lattice softening (decreasing in phonon frequency) [5], which means the varying Sn content leads to the spontaneous cubic-tetragonal phase transformation of A15 lattices. Recent experiment [6] also investigates the stress-induced transformation behaviors at low temperatures for polycrystalline Ti-51.8Ni (at.%) specimens. The alloying addition (Ti and/or Ta), which is introduced in most modern high-field Nb$_3$Sn conductors for increasing the electrical resistivity, suppressing the martensitic phase transformation and thus raising $B_{c2}$, also be approximately linear with the Sn content $\beta$ [7]. As for other additives, the ZrO$_2$ precipitates in Nb$_3$Sn wires could refine Nb$_3$Sn grain size such that change the pinning behavior [8]. Recent experiments also demonstrate a strong correlation between the composition and the superconducting properties in YBa$_2$Cu$_3$O$_{7-x}$ films [9].

For Nb$_3$Sn wires, the scaling law and the experiment show a disagreement in the field dependence of the pinning force at high reduced fields [10]. One of the reasons could be the inhomogeneity of microstructure and composition [10]. This also explains the observation that the scaling field lies below the experimental $B_{c2}$ of Nb$_3$Sn wires. In fact, the scaling field for the critical current reflects the average properties over the wire volume; it thus relates to the local variation of the critical field dependent on the composition gradient [11]. Cooley and the coauthors simulate the effect of Sn composition gradients on the superconducting properties of powder-in-tube (PIT) Nb$_3$Sn strand, with an ideal structure modelled by concentric shells with varying Sn content [7]. They found that different Sn profiles have a pronounced effect on the irreversibility fields defined by the extrapolation of Kramer plots $H_K$, and also that Sn gradients reduce the elementary pinning force, $H_K$ and the critical current density $J_c$ [7].

The temperature dependence of the upper critical field $H_{c2}(T)$ in inhomogeneous Nb$_3$Sn conductors, as the field-temperature phase boundary, is comprehensively investigated by Godeke *et al.* [2]. It is concluded that, the complete field-temperature phase boundary can be described with the simplest form of the Maki-DeGennes (MDG) relation, and independent of compositional variation, measuring technique, criterion for the critical field and strain state [2].

Various experiments have recently been conducted to investigate the dependence of the superconductivity and magnetic properties of Nb$_3$Sn samples on Sn content and disorder [12-14]. The underlying physics for the superconducting properties variation with the A15 composition in Nb$_3$Sn is however still not very clear. The already-existing physical formulas for this dependence are insufficient for describing the state-of-art experimental results of binary Nb$_3$Sn samples and practical Nb$_3$Sn wires. As for practical engineering significance, describing the superconducting properties dependence on the A15 composition in theory will facilitate the understanding of the optimization for the critical current density $J_c$, since tin composition and possible additives are important for the very high $J_c$ now achieved in commercial strands [15, 16].

In this paper, we extend the Ginzburg-Landau-Abrikosov-Gor'kov (GLAG) theory to account for the A15 composition dependence of the superconducting properties in Nb$_3$Sn. In this theory, the occurrence of $H_{c2}$ is due to the breaking of orbital pair and the Pauli paramagnetic limiting. Two scattering mechanisms should be considered in type-II superconductors like Nb$_3$Sn: electron-transport scattering by impurity (disorder) and spin-orbit scattering. As for A15 Nb$_3$Sn, composition inhomogeneity and its deviation from stoichiometry may cause defect and site disorder in Nb$_3$Sn lattices [4], which contribute most to the scattering by impurity. Based on this physical picture, $H_{c2}$ is correlated to the superconductivity parameters (the coherence length, the London penetration depth and etc.) as well as the scattering characteristics (the mean free path of electron transporting). As for the strong-coupling superconductor like Nb$_3$Sn, one should include the correction for electron-phonon interaction to the weak-coupling BCS value. Since the microscopic parameters mentioned above cannot be determined directly, we correlate them to the transition temperature, the normal-state resistivity and the coefficient of electronic heat capacity. The three material parameters have been extensively measured as a function of tin content [3, 4, 17, 18]. In this manner, we can determine the superconducting properties variation with composition concentration. The following section will present the detail.

## 2. Upper critical field variation with composition concentration

### 2.1. Upper critical field at vicinity of superconducting transition temperature

The best quality Nb$_3$Sn samples, with highest transition temperature $T_c$'s and resistance ratios ($RRR = \rho(300K)/\rho(20K)$), have very narrow resistive transitions [18]. The transitions tend to broaden in high fields. Measured ternary PIT wire also exhibits a narrow transition at a wide range of fields and temperatures [2]. At the vicinity of $T_c$, the upper critical fields $H_{c2}(T)$ are thus nearly the same, and independent of the selected critical-state criterion. We are then allowed to determine $H_{c2}$ near $T_c$ uniquely. The resistivities $\rho$ near $T_c$ are approximated as $\rho(T_c)$, measured at temperatures just above $T_c$, and the coefficient of electronic heat capacity $\gamma$ remains a constant which satisfies the low-temperature heat capacity formula without undergoing a specific heat jump.

In the microscopic physical concept, the scattering of impurity (Appendix A and Fig. A1) enters into the superconductivity of type II superconductors by changing the Ginzburg-Landau (GL) parameter $\kappa$ at the vicinity of $T_c$. We have $\kappa$ in two limiting cases, $\kappa_{\text{clean}}(T) = \kappa(T_c)\chi_1(T)$ without scattering effect and $\kappa_{\text{dirty}}(T) = \kappa_{\text{dirty}}(T_c)\chi_2(T)$ relevant to scattering, where $\kappa_{\text{clean}}(T_c) \approx 0.96\delta_L(0)/\xi_0$ and $\kappa_{\text{dirty}}(T_c) = 0.72\delta_L(0)/l$ (see Appendix B for GLAG description of the superconductivity parameters). Here, $\delta_L(0)$ is the London penetration depth of the magnetic field at 0K. $\xi_0$ is the standard coherence length. $l$ is the mean free path of electron transporting. $\kappa_{\text{clean}}$ and $\kappa_{\text{dirty}}$ refer to the clean limit ($\xi_0 \ll l$) and dirty limit ($\xi_0 \gg l$), respectively. $\chi_1(T)$ and $\chi_2(T)$ represent the temperature dependence of $\kappa$ for the clean limit and dirty limit, respectively; calculations show that they vary little with $T$ near $T_c$, $\chi_1(T) \approx \chi_2(T) \approx 1$ [19].

In light of the classic proposal [18, 20], the superconductivity parameters $\delta_L(0)$, $\xi_0$ and the scattering parameter $l$ involved in the GL parameter $\kappa$ can be linked to three independent material parameters, the transport scattering resistivity $\rho$, the low-temperature coefficient of electronic heat capacity $\gamma$ and the superconducting transition temperature $T_c$. Assuming a spherical Fermi surface and isotropic metal, we are allowed to use the electron conduction formula Eq. (A1) and the electron heat capacity relation Eq. (C1) in the GLAG description of the superconductivity parameters (Appendix C), and then express $l$, $\xi_0$, $\delta_L(0)$, $\kappa_{clean}$ and $\kappa_{dirty}$ as a function of $\rho$, $\gamma$ and $T_c$, Eqs. (C15)-(C19). Including the correction for the anisotropy in Nb$_3$Sn (Appendix D), one must consider the change of Fermi surface shape from the isotropic model, which leads to the corrected expressions, Eqs. (D3)-(D6).

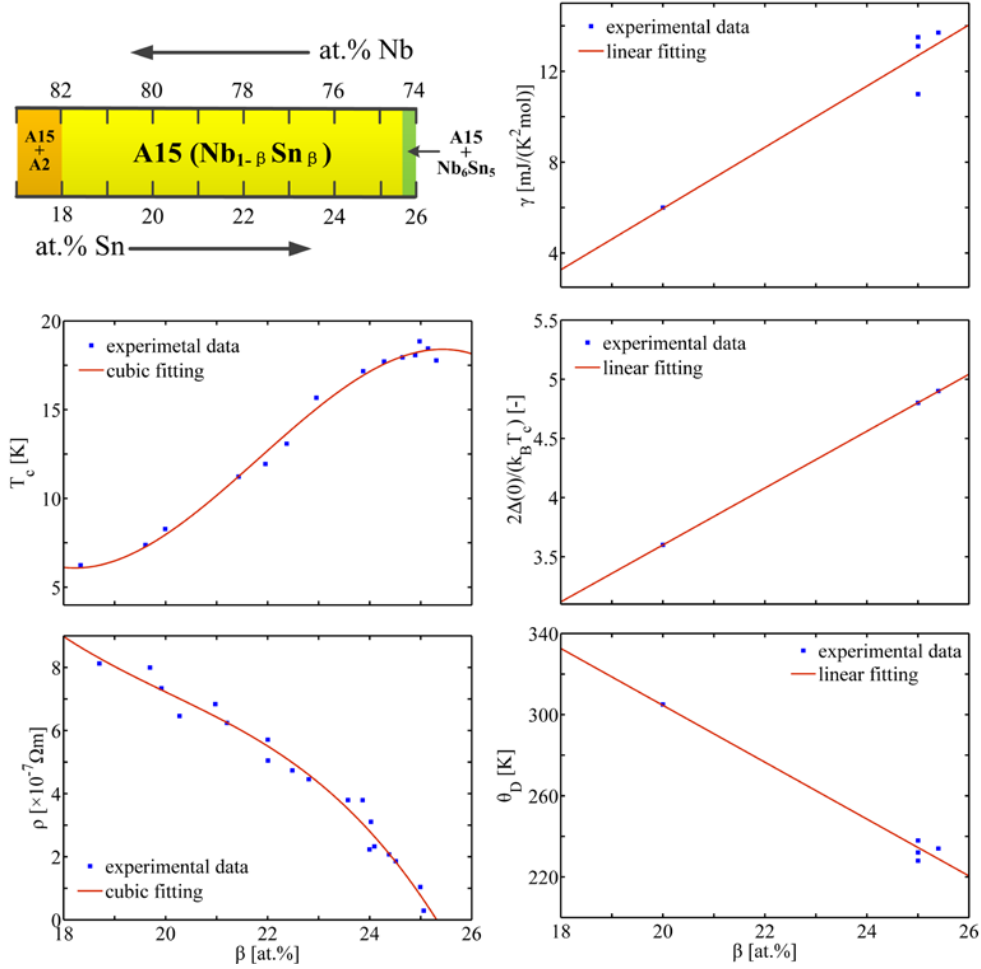

FIG. 1. Variations of material parameters $T_c$, $\rho$, $\gamma$, $\theta_D$ and $2\Delta(0)/(k_B T_c)$ with Sn content $\beta$ in the A15 range of binary Nb$_{1-\beta}$Sn$_\beta$. The experimental data are extracted from [3]. The solid lines represent the global fits to the experimental data in the A15 range.

Let us consider the dependences of $T_c$, $\rho$ and $\gamma$ with Sn content $\beta$. Experiments [3] show that $T_c$ variation with $\beta$ is essentially linear up to ~24 at%; while between 24 and 25 at% Sn content, $T_c$ versus $\beta$ has a saturation, Fig. 1. The A15 type lattice undergoes a spontaneous cubic-tetragonal transformation in this range. The lattice softening, as one of the consequences, implies a decreasing in the lattice stiffness ($\hbar\omega_D = k_B\theta_D$) such that it reduces $T_c$ according to McMillan $T_c$ equation [Eq. (6)]. The normal-state resistivity $\rho$, as measured just above $T_c$, decreases moderately with increase of $\beta$ within 18~24 at%. Approaching $\beta = 25$ at%, $\rho(\beta)$ exhibits an obviously stronger decrease. Below 24 at%, the coefficient of electronic heat capacity, $\gamma$, changes linearly with $\beta$ by $\sim 1.5 \text{mJ} \cdot \text{K}^{-2} \cdot \text{mol}^{-1}$ per at% Sn. Note that resistivity measurements [21] also show pressure-induced resistivity saturation in Fe17wt%Si. Following the treatment to the experimental data in [3], we use global fits to the experimental $T_c$, $\gamma$ and $\rho$ versus $\beta$ in the entire A15 range (Fig. 1 and Table 1),

$$T_c, \gamma \text{ or } \rho = \begin{cases} a_1 \beta + b_1, & \text{linear fit,} \\ a_2 \beta^3 + b_2 \beta^2 + c_2 \beta + d_2, & \text{cubic fit.} \end{cases} \quad (1)$$

TABLE 1. Material parameters variation with Sn content $\beta$ [at.%].

| | Linear fitting | ($\chi = a_1 \beta + b_1$) | | |
|---|---|---|---|---|
| $\chi$ | $a_1$ | $b_1$ | | |
| $\gamma$ | 1.3473 | -20.983 | | |
| $\theta_D$ | -14.018 | 584.94 | | |
| $2\Delta(0)/(k_B T_c)$ | 0.24043 | -1.2087 | | |
| | Cubic fitting | ($\chi = a_2 \beta^3 + b_2 \beta^2 + c_2 \beta + d_2$) | | |
| $\chi$ | $a_2$ | $b_2$ | $c_2$ | $d_2$ |
| $T_c$ | -0.065923 | 4.3157 | -91.612 | 641.31 |
| $\rho$ | -0.021449 | 1.2923 | -26.735 | 196.6 |

We are now ready to calculate the dependence of the superconducting characteristic lengths $\delta_L(0)$, $\xi_0$ and $l$ with the Sn concentration $\beta$, by substituting Eq. (1) into Eqs. (D4), (D5) and (D6). The London penetration depth $\delta_L(0)$ increases with Sn content, while the coherence length $\xi_0$ is reduced, Fig. 2. $\delta_L(0)$ and $\xi_0$ are roughly with the same order of magnitude over the A15 range. By comparing the coherence length $\xi_0$ and the electronic mean free path $l$, one may distinguish the clean limit, the dirty limit and the intermediate state at any A15 composition. At lower Sn content, $\xi_0$ maintains rather high value compared to $l$; but this difference is mitigated as $\xi_0$ continues to decrease and $l$ increase, for rising $\beta$ until the phase transformation boundary. In the tetragonal phase range, $\xi_0$ is approximately equal to or even lower than $l$. This implies that, as Sn content gradually approaches the stoichiometry, Nb$_3$Sn undergoes a transition from the "dirty" limit ($\xi_0 \gg l$) to the "clean" limit ($\xi_0 \ll l$), and the phase transformation boundary may be taken as the boundary of this transition. In Fig. 2, we also presents the variation of GL parameter $\kappa$ with tin content, calculated by Eqs. (C19) and (D3). $\kappa$ in the two limiting cases, $\kappa_{\text{clean}}$ and $\kappa_{\text{dirty}}$, have the opposite change with Sn content; their change nearly counteract one another in the cubic phase range. While in the tetragonal phase range, $\kappa_{\text{dirty}}$ exhibits a more severe decrease compared to the increase in $\kappa_{\text{clean}}$. Thus, $\kappa = \kappa_{\text{clean}} + \kappa_{\text{dirty}}$ varies little in the cubic phase range, but decreases largely in the tetragonal phase.

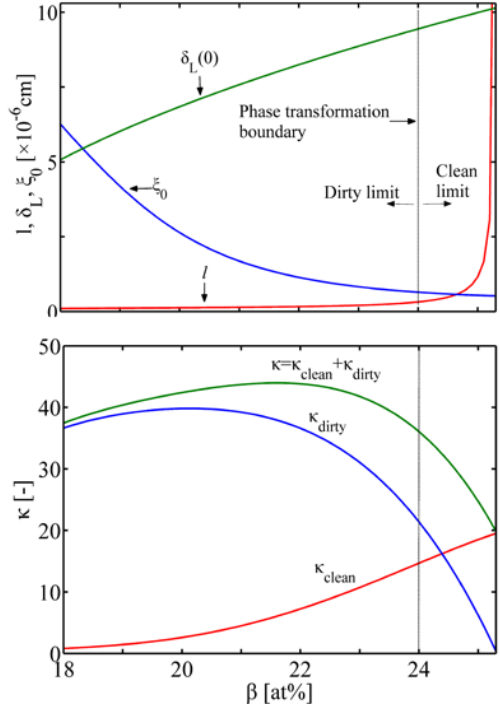

FIG. 2. Composition dependence of the London penetration depth $\delta_L(0)$, coherence length $\xi_0$ and electronic mean free path $l$ (Top), and the GL parameter $\kappa$ (Bottom).

The magnetic properties of type II superconductors gives $H_{c2} = \sqrt{2}\kappa H_c$ at the vicinity of $T_c$, where $H_c$ is the thermodynamic critical field [22]. Combining this with the expressions $\kappa$ and $H_c$, Eqs. (C19), (C20) and (D3), renders the upper critical field $H_{c2}$ in the clean limit and the dirty limit, Eqs. (C14) and (D7) (see Appendix C and D for detail). Thus, $H_{c2}$ at the intermediate $\xi_0/l$ can be obtained by summing $H_{c2,\text{clean}}$ and $H_{c2,\text{dirty}}$,

$$H_{c2} = C_1 \chi_1(T) T_c^2 (1-T/T_c) \gamma_{mJ}^2 + C_2 \chi_2(T) T_c (1-T/T_c) \gamma_{mJ} \rho_{\Omega m}. \tag{2}$$

Here, $C_1 = 1.979 \times 10^8 \pi^{1/3} ce^{-1} k_B^{-2} \hbar (n^{2/3} S/S_F)^{-2}$ and $C_2 = 1.356 \times 10^{-6} \pi^{-1} ce k_B^{-1}$ in which the physical constants are listed in Table A1. The involved three material parameters $T_c$, $\gamma_{mJ}$ and $\rho_{\Omega m}$ as well as their variations with the composition concentration in A15 Nb$_3$Sn have been widely measured [3], Eq. (1) and Fig. 1. Thus, substituting Eq. (1) into Eq. (2), one finds the dependence of the upper critical field $H_{c2}$ with Sn content. Note that the above formulas are valid at the vicinity of the superconducting transition temperature $T_c$.

### 2.2. Corrections to upper critical field for electron-phonon interaction

A15 type Nb$_3$Sn is a strong-coupling superconductor such that the corrections for electron-phonon (EP) interaction are required. We obtain the upper critical field, Eq. (2), based on the breaking and scattering of cooper pairs in a weak-coupled interaction. Equation (2) is corrected for EP interaction in two ways: renormalizing the material parameters, and introducing correction parameters to superconductivity itself [18, 23, 24].

We determine which material parameter should be renormalized in terms of Grimvall principle [23]. Both the density of states and the wave function are renormalized by EP interaction, and the electron mass in the absence of EP interactions is replaced by a renormalized electron mass $m = m^b(1+\lambda_{ep})$, where b refers to band values. The electronic density of states $\upsilon(\mu)$ at Fermi level is renormalized by an enhancement factor $1+\lambda_{ep}$ where $\lambda_{ep}$ is the EP interaction parameter, and thus the coefficient of the electronic heat capacity $\gamma_{mJ}$ is enhanced by $(1+\lambda_{ep})\gamma_{mJ}$. However, $\gamma_{mJ}$ is not changed for EP renormalization in Nb$_3$Sn, since there are no EP renormalization effects in the change of the Fermi surface dimensions on alloy and the change in $\upsilon(\mu)$ always depends on the Fermi level and follows almost rigidly any shift in Fermi energy. The electrical resistivity $\rho_{\Omega m}$ [Eq. (A1)] is not renormalized, since the renormalization of the electron mass $m$ exactly cancels against the renormalization of the scattering matrix element as it enters the averaged time $\tau$ between collisions. We find that, these are implicitly followed by Devantay $at$ $el.$ [25], who do not take renormalizations on $\gamma_{mJ}$ and $\rho_{\Omega m}$. Thus, we take the only EP correction in Nb$_3$Sn by multiplying a factor $\eta_{H_{c2}}(T)$ for $H_{c2}$,

$$H_{c2} = \eta_{H_{c2}}(T)[C_1\chi_1(T)T_c^2(1-T/T_c)\gamma_{mJ}^2 + C_2\chi_2(T)T_c(1-T/T_c)\gamma_{mJ}\rho_{\Omega m}]. \tag{3}$$

Here, $\eta_{H_{c2}}(T)$ is the ratio of the strong-coupled magnetic pair-breaking parameter to the weak-coupled BCS value, and can be evaluated by the detailed EP spectrum [26].

If $T_c$ and the energy gap $\Delta(0)$ at 0K are determined experimentally, we have the strong-coupling correction $\eta_{\Delta(0)}$ to $\Delta(0)$, $\eta_{\Delta(0)} = (2\Delta(0)/k_BT_c)_{meas}/(2\Delta(0)/k_BT_c)_{BCS} = (2\pi/\eta)^{-1}(2\Delta(0)/k_BT_c)_{meas} \approx 0.283(2\Delta(0)/k_BT_c)_{meas}$ in which the BCS equation $\Delta(0) = (\pi/\eta)k_BT_c$ with $\eta = 1.78$ has been used for the second equal sign. The characteristic (equivalent Einstein) frequency $\omega_0$ [erg] is then determined using [18] (Appendix E)

$$\eta_{\Delta(0)} = 1 + 5.3(\omega_0 k_B^{-1}T_c^{-1})^{-2}\ln(\omega_0 k_B^{-1}T_c^{-1}). \tag{4}$$

The EP correction to $H_{c2}$, $\eta_{H_{c2}}(T_c)$, can thus be obtained,

$$\eta_{H_{c2}}(T_c) = 1 + \pi^2(\omega_0 k_B^{-1}T_c^{-1})^{-2}[0.6\ln(\omega_0 k_B^{-1}T_c^{-1}) - 0.26]. \tag{5}$$

The another correction parameter (not involved in EP correction for Nb$_3$Sn), the EP interaction parameter $\lambda_{ep}$, is determined by the McMillan strong-coupled $T_c$ equation [24]

$$T_c = \frac{\theta_D}{1.45}\exp[-\frac{1.04(1+\lambda_{ep})}{\lambda_{ep} - \mu^*(1+0.62\lambda_{ep})}], \tag{6}$$

where $\theta_D$ is the Debye temperature and $\mu^*$ is the pseudo potential parameter for electron Coulomb repulsion (for Nb$_3$Sn $\mu^* \approx 0.2$). Note that to obtain a more accurate $\lambda_{ep}$ for Nb$_3$Sn one should use the Allen-Dynes $T_c$ formula, which extends the application range $0 < \lambda_{ep} < 1.5$ of Eq. (6) to a large $\lambda_{ep}$ value. Experiments give $\lambda_{ep} \approx 1.8$ for Nb$_3$Sn. However, the available experimental data is $\theta_D$ and not the characteristic phonon frequency required in the Allen-Dynes $T_c$ formula. Thus we still apply Eq. (6) at the cost of some accuracy.

The corrections for EP interaction require the information of the Debye temperature $\theta_D$ and the energy gap $\Delta(0)$ at 0K, Eq. (3). Unfortunately, the available data are reported at very limited Sn content. We use linear fits to $\theta_D$ and $(2\Delta(0)/k_B T_c)$ according to the identification [3] of the gradual changes from strong coupling to weak coupling with decreasing $\beta$, Fig. 1.

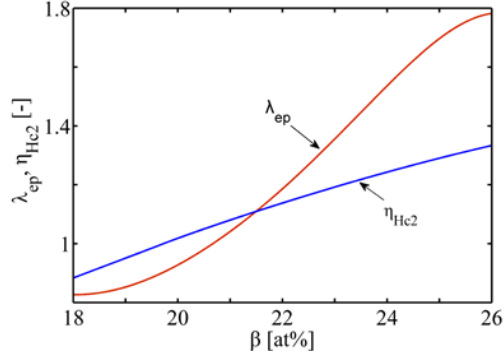

FIG. 3. EP correction parameters, the Debye temperature $\theta_D$ and energy gap $\Delta(0)$, versus tin content.

Figure 3 presents the calculated EP correction parameters in terms of Eqs. (1), (5) and (6). The result of the strong-coupling correction $\eta_{H_{c2}}$ to $H_{c2}$ approximately equals to the calculated value [18] of 1.17 at $T_c = 17.8\text{K}$. The calculated EP interaction parameter $\lambda_{ep}$ at the stoichiometry 25at% Sn is slightly lower than the generally accepted value of 1.8 [27]. However, at larger derivation from the stoichiometry, the calculation differs relatively larger from the value of 1.8.

**2.3. Upper critical field at temperature 0K**

In the following we will concern the behavior of $H_{c2}$ at temperatures far from $T_c$, where the GLAG theory does not apply. This will generalize the above results near $T_c$ [e.g. Eq. (3)] to a wide temperature range and up to 0K. The temperature dependence of $H_{c2}$ in the scaling law for Nb$_3$Sn is determined by the Maki-de Gennes (MDG) relation [28],

$$H_{c2}(T) = H_{c2}(0)[1-(T/T_c)^{\lambda}]. \tag{7}$$

Recent measurements suggest $\lambda \approx 1.52$; this value has a universe applicability to a wide range of off-stoichiometric samples and different methods of determining $H_{c2}$ [11]. This value of $\lambda$ is also the power determined from the MDG theory and Eliashberg theory [28, 29]. Godeke *et al.* demonstrate that the MDG description (7) is universal for $H_{c2}(T)$ relation of Nb$_3$Sn independent of the compositional variation [2, 11]. Recent experiment shows that this relation is applicable for Nb$_3$Sn with or without undergoing the cubic-to-tetragonal transition [12]. We are then allowed to extrapolate $H_{c2}$ at vicinity of $T_c$ to $H_{c2}(0)$ at temperature 0K for any Sn content. If $H_{c2}(T_0)$ with $T_0 \to T_c$ has been obtained using Eq. (3), then one can deduce $H_{c2}(0)$ as

$$H_{c2}(0) = H_{c2}(T_0)/[1-(T_0/T_c)^{\lambda}]. \qquad (8)$$

We assume that EP interaction correction to $H_{c2}$ far from $T_c$ is the same as that for $H_{c2}$ near $T_c$.

WHH equation, derived from Gor'kov superconductivity theory (Green function method) and taking into account electron spin and spin-orbital scattering [30], is capable of giving rather satisfactory descriptions for $H_{c2}$ behavior of a wide range of commercial and experimental Nb$_3$Sn wires. This equation can be written as a simple form

$$H_{c2}(0) = 0.69 H_{c2}' T_c, \qquad (9)$$

where $H_{c2}' = -(dH_{c2}/dT)_{T_c}$ [14]. Substituting the derivative of Eq. (3) into Eq. (9) and then comparing to Eq. (2), one finds $H_{c2}(0) = 0.69 H_{c2}(T_0)/(1-T_0/T_c)$. It is further shown that the only difference between WHH equation and MDG relation is the power of $(T_0/T_c)$; Since $T_0/T_c \to 1$ this difference has very limited impact on the $H_{c2}(0)$ values. This is consistent with the viewpoint [30, 31] that MDG relation is a good approximation to Werthamer theory.

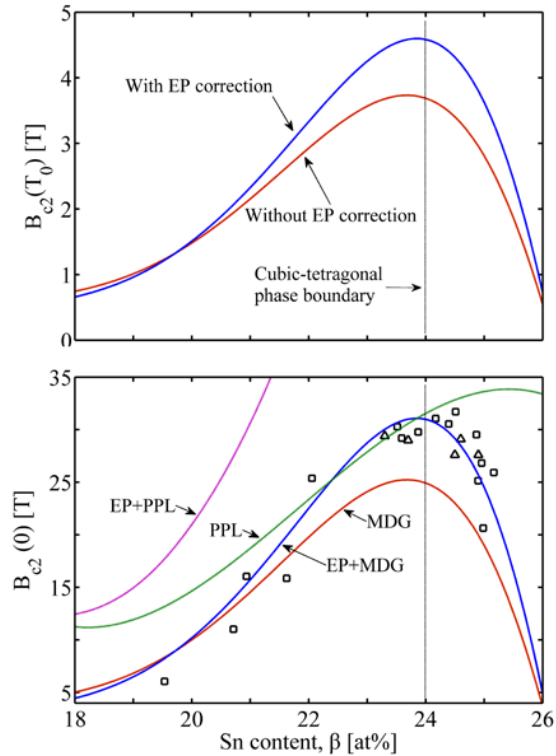

FIG. 4. Upper critical field $B_{c2}$ variation as a function of tin content $\beta$ at: the vicinity of $T_c$ (Top) and 0K temperature (Bottom). MDG: extrapolation of $H_{c2}(T_c)$ with MDG relation, Eqs. (1), (2) and (8); EP+MDG: extrapolation of EP corrected $H_{c2}(T_c)$ with MDG, Eqs. (1), (2), (3) and (8); PPL: Pauli paramagnetic limit, Eqs. (1) and (10); EP+PPL: EP corrected Pauli paramagnetic limit, Eqs. (1) and (11); □: Experiment dataset 1 [3]; △: Experiment dataset 2 [12].

We now calculate the upper critical field $B_{c2}$ near $T_c$ using Eqs. (1), (2) and (3), Fig. 4. The temperature $T_0$ near $T_c$ is designated as $T_0 = 0.9T_c$. One finds the dependence of $B_{c2}(T_0)$ on Sn content $\beta$ with/without the EP correction. The results of $B_{c2}(T_0)$ are then substituted into the MDG relation, Eq. (8), to obtain $B_{c2}$ at 0K, $B_{c2}(0)$. Since the WHH relation is equivalent to the MDG relation, we do not implement it redundantly. The calculation results are impressive (Fig. 4): $B_{c2}(0)$ obtained by $B_{c2}(T_0)$ with EP correction and then extrapolated by MDG relation (EP+MDG) is in good agreement with the experiments. This validates the above GLAG descriptions, and also the MDG description for the temperature dependence of the upper critical field at any composition over the A15 phase field.

We find that, the cubic-tetragonal phase boundary at ~24at% Sn separates the increasing $B_{c2}$ versus $\beta$ from the decreasing $B_{c2}$ versus $\beta$. As shown both in the experimental $B_{c2}(0)$ versus tin content and the EP+MDG curve, the maximum ~29T of $B_{c2}(0)$ appears at ~24at% Sn, and at both sides of the peak, $B_{c2}(0)$ decreases as tin content deviates more from 24at% Sn. In the tetragonal phase, $B_{c2}(0)$ has a more serious reduction; however, at the vicinity of 24at% Sn, $B_{c2}(0)$ have nearly the same values at both sides of 24at%Sn, namely in the cubic phase range and the tetragonal phase range. This is consistent with the experiment by Zhou *et al.* [12], their tetragonal phase [ $B_{c2}(0.3K) = 29.1T$ at $\beta = 24.6 \pm 0.2$ at% ] and cubic phase [ $B_{c2}(0.3K) = 29.0T$ at $\beta = 23.7 \pm 0.4$ at% ] samples exhibiting almost identical $B_{c2}(0) \approx 29 \pm 0.2$ T. We infer that, this coincidence occurs in a limited range, where the tin content deviates small from the phase transformation boundary; for a large deviation there is a stronger depression of $B_{c2}(0)$ in the tetragonal phase. This phenomenon is caused by the underlying relationship between the superconductivity of Nb$_3$Sn and the related material parameters; physical properties of the latter is continually changed by the spontaneous cubic-tetragonal transformation.

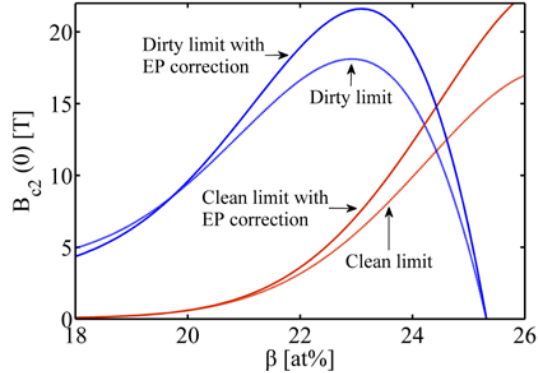

FIG. 5. Upper critical field $B_{c2}(0)$ in the clean and dirty limit case.

We now consider the upper critical field $B_{c2}(0)$ decomposing into the component in the clean limit case [Eqs. (D7)] and that in the dirty limit case [Eqs. (C14)], $B_{c2}(0) = B_{c2,\,\text{clean}} + B_{c2,\,\text{dirty}}$. One finds that $B_{c2,\,\text{clean}}$ increases as $\beta$ raised, but for contrast the dirty one increases up to 24 at% and then deceases drastically, Fig. 5. As we already know, Nb$_3$Sn undergoes a transition from the "dirty" limit to "clean" limit as $\beta$ increased. By comparing Figs. 5 and 6, we find that, within 18at%~24at%, the component $B_{c2,\,\text{dirty}}$ dominates in $B_{c2}(0)$ and determines the trend of $B_{c2}(0)$ curve. Once crossing over the phase transformation boundary, $B_{c2,\,\text{clean}}$ takes over $B_{c2}(0)$ variation with $\beta$.

### 2.4. Limit for upper critical field: Pauli paramagnetic limit

Preferential Pauli-paramagnetic lowering of normal-state free energy should place a limit on the orbital-pair-breaking $H_{c2}$ of filamentary high-field superconductor [32],

$$H_{c2}(0) \leq H_p(0) \equiv 1.84 \times 10^4 T_c \text{ [Oe]}. \tag{10}$$

This equation is a rough estimation for the limit field $H_p(0)$, since the assumption that the zero temperature difference between superconducting and normal-state magnetizations be at least equal to the Pauli conduction-electron-spin magnetization, is somewhat contrary to experimental results [33]. The correction to $H_p(0)$ for EP interaction is given by

$$H_p(0) = 1.84 \times 10^4 \eta_{H_c}(0)(1+\lambda_{ep})^{1/2} T_c \text{ [Oe]}, \tag{11}$$

where $\eta_{H_c}(T_c) = 1 + \pi^2 (\omega_0 k_B^{-1} T_c^{-1})^{-2}[1.1\ln(\omega_0 k_B^{-1} T_c^{-1}) + 0.14]$ assuming $\eta_{H_c}(T)$ is independent with temperature [18].

In Fig. 4, PPL (Pauli paramagnetic limit), Eqs. (1) and (10), slightly lowers the EP+MDG curve after ~23at% Sn, and it thus provides a good boundary with the experimental data before the phase transformation. This justifies that, PPL is independent with the EP interaction in Nb$_3$Sn. In Fig. 4, EP corrected PPL [Eqs. (1) and (11)] is much higher than the experimental curve and has no restriction to $B_{c2}(0)$. This is consistent with the view by Orlando *et al.* [18], who demonstrate that EP corrected PPL has nothing to do with Nb$_3$Sn superconductivity, since the strong EP interaction in Nb$_3$Sn increases largely the Pauli limiting field above its BCS value and the spin-orbit scattering is less involved. This explains why we only take the impurity scattering into account while exclude the spin-orbit scattering.

### 3. Temperature dependence of upper critical field

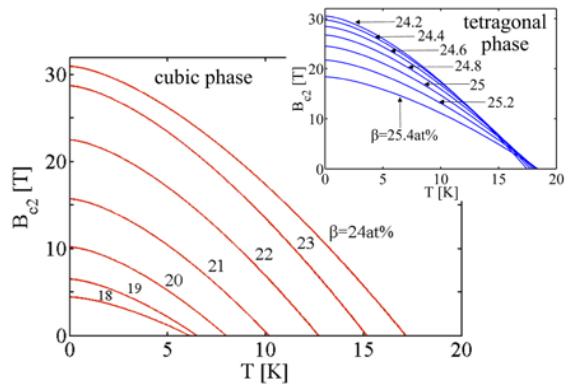

FIG. 6. Temperature dependence $B_{c2}(T)$ at varying Sn content in the range of the cubic phase at increment of 1at% from 18at% to 24at% Sn. The insert figure represents the situation in the tetragonal phase for 0.2at% increment between 24.2at% and 25.4at% Sn.

TABLE 2. Calculated *H-T* phase boundary and comparison to experiments.

| Samples | $\beta$[at%] | $\rho(T_c)$[$\mu\Omega\cdot$cm] | $\gamma$[mJ/molK$^2$] | $T_c$[K] | $B_{c2}(0)$[T] |
|---|---|---|---|---|---|
| Thin films by Orlando *et al.* [18] | -/23.59 [a] | 35/35.0 | -/10.8 | 16.0/16.4 | 29.5/30.8 |
|  | -/24.94 | 9/9.1 | -/12.6 | 17.4/18.2 | 26.3/25.2 |
| Bulk by Godeke *et al.* [2] | -/24.33 | 22(50%)/22.0 | -/11.8 | 16.5/17.6 | 27.4/30.1 |
|  | -/24.33 | 22(90%)/22.0 | -/11.8 | 16.6/17.6 | 28.3/30.1 |
| Polycrystal by Guritanu *et al.* [17] | -/24.46 | -/19.4 | 13.7/11.8 | 17.8/17.8 | 25.0/29.4 |

[a] Before the oblique line is the experimental value and after that presents the calculated value.

Now, we focus on the temperature dependence of $B_{c2}$ at any A15 composition concentration, Eqs. (1), (3), (7) and (8). In Fig. 6, the temperature dependence $B_{c2}(T)$ increases at any temperature as raising the Sn content $\beta$ within the cubic phase range. The tetragonal phase exhibits a reverse behavior, $B_{c2}$ at 0K and at most of other temperatures increasing with reduced Sn content. Orlando *et al.* show that in thin films $B_{c2}(0)$ is increased with $\rho$ rising, however $T_c$ is suppressed [2, 18]. For $\rho(T_c) = 35\mu\Omega\cdot$cm, there exists $B_{c2}(0) = 29.5$T and $T_c = 16.0$K; while $\rho(T_c) = 9\mu\Omega\cdot$cm leads to $B_{c2}(0) = 26.3$T and $T_c = 17.4$K [18]. This corresponds to the calculated composition dependence in the tetragonal phase range, Figs. 1 and 6. The measured bulk needle by Godeke *et al.* exhibits a similar behavior: $\rho(T_c) = 22\mu\Omega\cdot$cm at a 50% normal-state resistance criterion corresponds to $B_{c2}(0) = 27.4$T and $T_c = 16.5$K, and at a 90% criterion $B_{c2}(0) = 28.3$T and $T_c = 16.6$K [2]. A summary of the calculations and the comparisons to experiments is given in Table 2. Through the theoretical expressions, we also figure out the material parameter that is undetermined by experiments, Table 2.

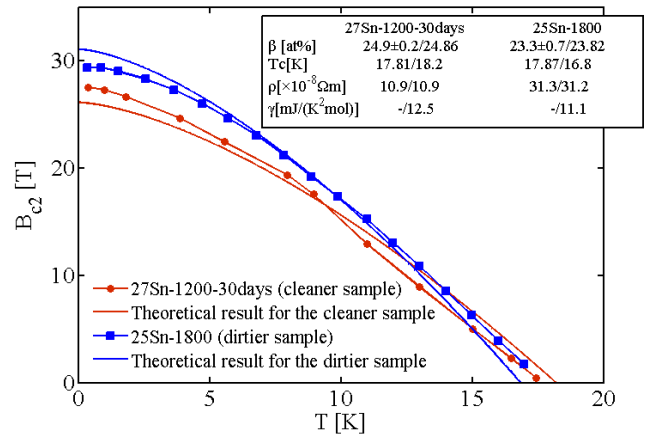

FIG. 7. Temperature dependence $B_{c2}(T)$ for two homogenous off-stoichiometric Nb$_3$Sn samples, and comparison to experiments [12].

We also calculate $B_{c2}(T)$ for two rather highly homogenous Nb$_3$Sn samples, Fig. 7. Their material parameters, reproduced from [12], are listed in the insert of the figure. Note that in the absence of the raw experimental value of $\gamma$, we infer it from Eq. (1). Calculations show that, the dirtier sample (25Sn-1800) with the larger resistivity ($31.3\mu\Omega\cdot cm$) always exhibits a higher $B_{c2}$ at any temperature while $T_c$ is suppressed. This follows the above observation by Orlando *et al.* [18] and Godeke *et al.* [2]. The little discrepancy from the experiment may be attributed to the absence of the raw data of $\gamma$ and the greater uncertainty in experimentally determining the composition concentration (fluctuation of ±0.7at% Sn for the dirtier sample compared to ±0.2at% for the cleaner one).

## 4. Flux pinning force and Kramer plot

We now extend the method for the upper critical field $B_{c2}$ to account for the pinning behavior at different composition concentrations. Using the flux pinning model proposed by Kramer [34], the pinning force per volume for Nb$_3$Sn conductors, $F_p(B)$, is given by [11]

$$F_p(B) = J_c B = 12.8\kappa^{-2} B_{c2}^{2.5} (B/B_{c2})^{0.5} (1 - B/B_{c2})^2 \ [GN\cdot m^{-3}], \quad (12)$$

where $J_c$ is the critical current density. $F_p(B)$ is associated with Sn concentration through the composition dependences $B_{c2}(T,\beta)$ and $\kappa(\beta)$, which are formulated by Eqs. (1), (3), (7), (8), (C19) and (D3). The Kramer function $f_K(B) = J_c^{0.5} B^{0.25}$ is linear with the magnetic induction $B$ and can identify $B_{c2}$ at which $f_K(B) = 0$ [11],

$$f_K(B) = J_c^{0.5} B^{0.25} = 1.1\times 10^5 \kappa^{-1} (B_{c2} - B) \ [A^{0.5} m^{-1} T^{0.25}]. \quad (13)$$

From Fig. 8 we observe a profound influence of the composition concentration on the field dependent pinning force $F_p(B)$. Within the cubic phase range, the increase of Sn content raises the $F_p(B)$ remarkably and shifts the peak in each $F_p(B)$ curve to the high field region. This is associated with the composition dependent superconducting properties. To be specific, the magnitude of $B_{c2}$ determines the position of the peak in $F_p(B)$, while the height of $F_p(B)$ is related to both $B_{c2}$ and $\kappa$, Eq. (12). $B_{c2}$ in the cubic phase range increases as $\beta$ rising (Fig. 4) while $\kappa$ varies little (Fig. 2), thus resulting in the shift of the peak and the increase in the height of $F_p(B)$. This case differs from that for the tetragonal phase range, where rising $\beta$ leads to a drastic decrease in both $B_{c2}$ and $\kappa$. As in Fig. 8, the peak shifts to the left and the height is nearly the same (due to $F_p(B) \propto \kappa^{-2} B_{c2}^{2.5}$). The experimental data of a Nb$_3$Sn conductor locate roughly between the calculated $F_p(B)$ curves for different Sn contents. This implies that, to describe a real Nb$_3$Sn conductor one may consider the composition gradient in the conductor which results in a weighted average of local homogenous composition properties. For Kramer plot, we find its variation consistent with $F_p(B)$ for the similar reasons, Fig. 8.

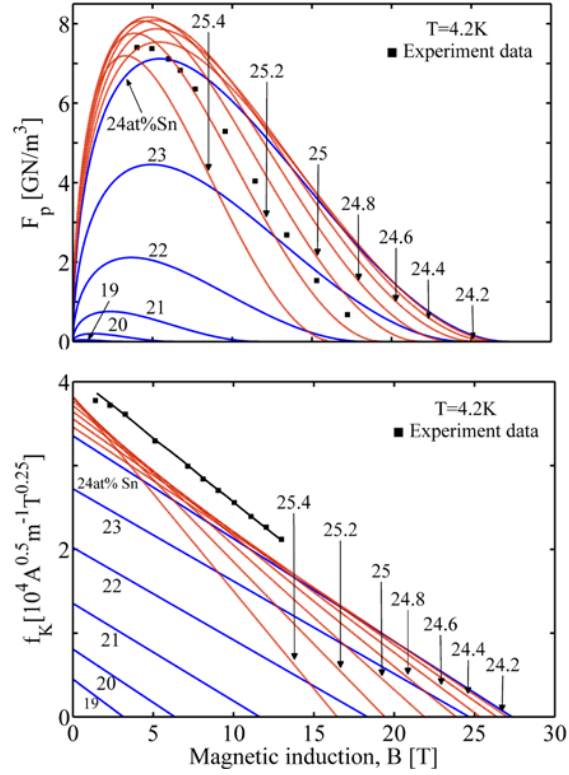

FIG. 8. Effect of composition concentration on the magnetic field dependence of the flux pinning force (Top) and Kramer plot (Bottom). The experiment data are extracted from [28] [11].

## 5. Composition gradient effect on practical Nb$_3$Sn wires

In the following we will apply our descriptions for the superconducting properties variation with tin content in the concentric shells model proposed by Cooley *et al.* [7], to arrive at the real composition dependent behavior in practical Nb$_3$Sn conductors. In the Nb$_3$Sn filament of PIT wires, the A15 layer exists between a central Sn-rich core and a coaxial Nb tube, Fig. 9. This structure is convenient for using a series of concentric shells with varying Sn concentration to simulate the composition inhomogeneity in the wire. The Sn content $\beta$ varies with the position in the A15 layer [7],

$$\beta = 18 + 3.5[1 - r^N + (1-r)^{1/N}] \, [\text{at\%}], \tag{14}$$

where $N$ indicates the severity and steepness of the overall gradient and $r$ is the normalized radius of the filament cross section. The radius $r$ is set to be 0 at the A15/Sn interface and reaches its maximum 1 at the Nb/A15 interface. The actual position in this area is counted as $R = 10 + 5r$ [μm], indicating that the inner radius of the A15 layer is $R_{\min} = 10\,\mu\text{m}$ and the outer radius is $R_{\max} = 15\,\mu\text{m}$.

For PIT wires, the prefactor of the flux pinning force differs from the Kramer model [Eq. (13)], $F_p(B) = J_c B = 0.35 B_{c2}^2 (B/B_{c2})^{0.5} (1 - B/B_{c2})^2$ [GN·m$^{-3}$] and $f_K(B) = 1.871 \times 10^4 B_{c2}^{-0.25} (B_{c2} - B)$ [A$^{0.5}$m$^{-1}$T$^{0.25}$]. We can then link $B_{c2}(T, \beta)$ and $J_c$ to the radius $r$, and the local magnetic moment is calculated as $m_o = \pi R^2 t L J_c$ [Am$^2$], where $t$ is the thickness of each of the 100 shells, i.e. $t = (R_{max} - R_{min})/100$ and $L$ is the sample length.

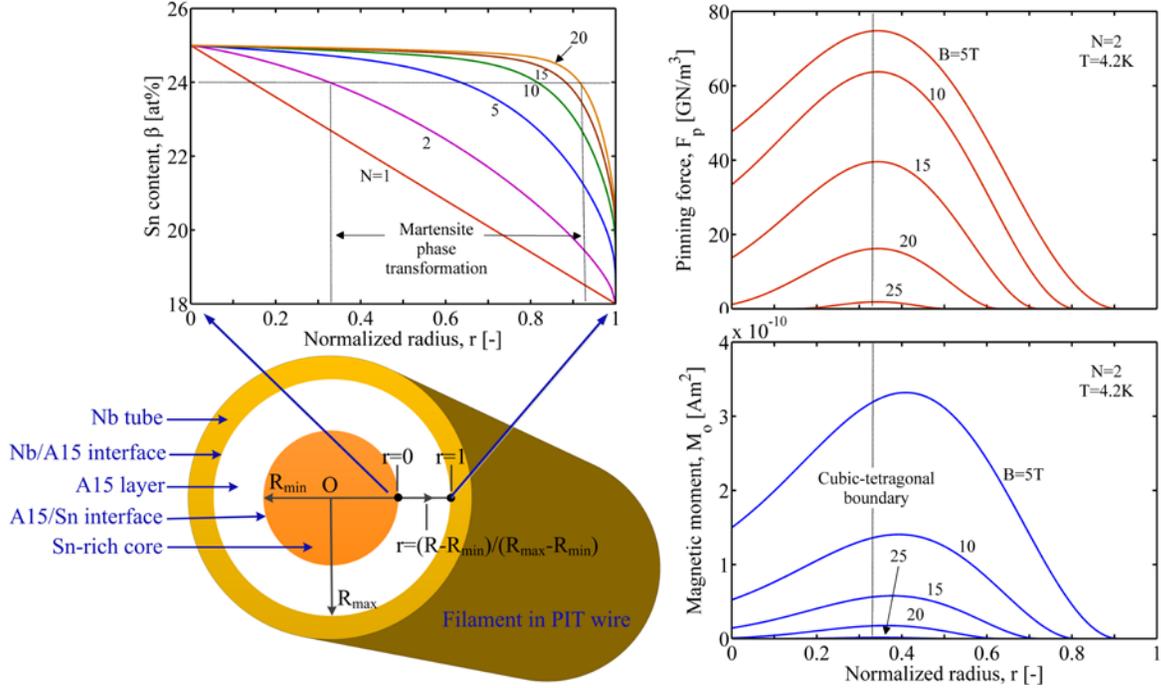

FIG. 9. Distribution of the Sn content $\beta$ (Top left), pinning force $F_p$ (Top right) and magnetic moment $m_o$ (Bottom right) on the cross section of PIT wire filament (Bottom left), at temperature $T$=4.2K. $N$ indicates the severity of the overall gradient.

Figure 9 presents the Sn content variation with the radius $r$ at different gradients $N$. It is found that, larger steepness of the overall gradient $N$ has a more drastic decrease near the outer radius (closer to the real situation). The distribution of the magnetic moment $m_o$ along the radius $r$ at different applied magnetic fields is also presented in Fig. 9. At any magnetic field $B$, the regions near the Nb/Nb$_3$Sn ($r=1$) and/or Nb$_3$Sn/Sn ($r=0$) interfaces appear no magnetic moment and thus loss of superconductivity. The larger region of the loss occurs for the higher magnetic field, with a suppression in the magnitude of magnetic moment. In fact, the magnetic moment is related to the local critical current density and thus the pinning force $F_p(B)$. One can find the Sn content dependence of $F_p(B)$ from Fig. 8; in the cubic phase range, $F_p(B)$ is raised with the Sn content increasing at any field, and $F_p$ for lower Sn content is more probably turns to disappear at higher fields. These are the underlying reasons for the vanishing of the magnetic moment. The tetragonal phase has a similar corresponding relationship. So, the loss of superconductivity near the boundary of A15 layer is mainly associated with the change of the flux-pinning behavior due to the A15 composition variation.

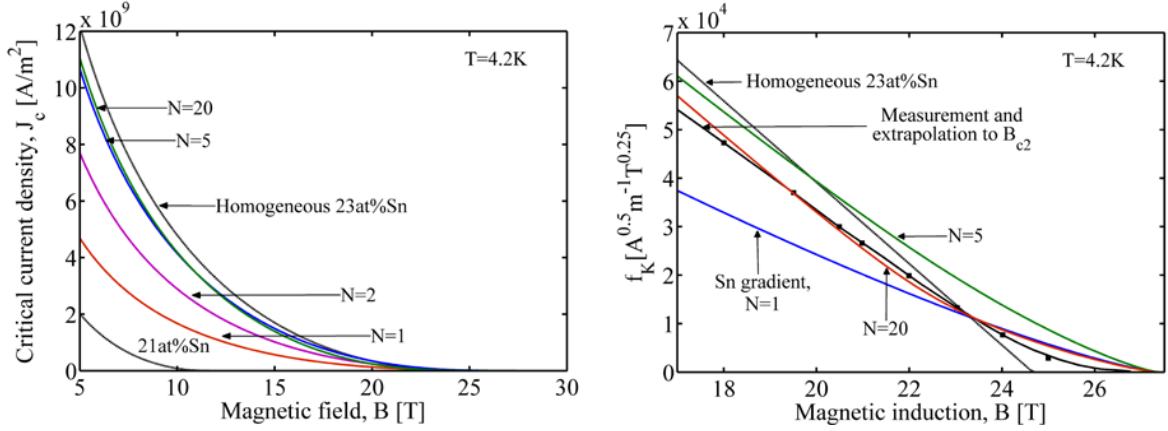

FIG. 10. Effect of the composition gradient on the critical current density (Left) and Kramer function $J_c^{0.5}B^{0.25}$ (Right) of PIT wires. The measurements on SMI ternary PIT wire are duplicated from [2]. The Kramer plot with a composition gradient $N = 20$ has a positive curvature approaching the experiment.

Summing the contributions $m_o$ from each shell results in the total magnetic moment $m_{ot}$, which can be measured by magnetometry. The critical current density $J_c$ of the wire is then expressed as $J_c = 3m_{ot}/[\pi L(R_{max}^3 - R_{min}^3)]$ [A/m$^2$] [7]. In Fig. 10, we present the variation of $J_c$ as a function of the magnetic field $B$ for different Sn gradients $N$. The homogenous sample results are presented to emphasize the effect of the tin gradient. The results of $J_c(B)$ allow one to depict the Kramer plot $f_k(B)$. The Kramer plot [2] for the measurements on a SMI ternary PIT wire exhibits an anomalous curvature "tail" at the vicinity of $B_{c2}$, Fig. 10. Two calculated homogenous samples with 21at% and 23at% appear no curvature, however, the inhomogeneous samples have a curved $f_k(B)$ when approaching $B_{c2}$. An acceptable agreement with the experiment is found for the conductor with gradient severity $N=20$. At this point, we quantitatively predicts the positive curvature in the Kramer plot of the PIT wire, emphasizing the importance of the composition inhomogeneity in the superconducting properties of practical Nb$_3$Sn conductors. The composition dependent pinning force and Kramer function explains the disagreement between the results of the scaling laws and the experiments for practical inhomogeneous conductors.

## 6. Conclusions

Although Nb$_3$Sn has been extensively used in fusion engineering area like ITER, the dependence of superconductivity with inhomogeneous composition is established incompletely in theory. Based on the GLAG theory frame, we derive a series of expressions for the superconductivity parameters and the upper critical field as a function of the three material parameters. These relations have a complete self-consistent theory basis describing the variation of superconductivity of Nb$_3$Sn with Sn content. The correction for EP interaction are included. The theoretical estimation of $H_{c2}(0)$ variation with Sn content, provided with the fits to the material parameters, shows an acceptable agreement with the experiments.

Nb$_3$Sn undergoes a transition from "dirty" limit ($\xi_0 \gg l$) to "clean" limit ($\xi_0 \ll l$) as Sn content gradually approaches the stoichiometry. The change in superconductivity at the vicinity of the critical temperature and in the related material parameters determines the composition dependence of the upper critical field. The MDG description is universal for describing *H-T* phase boundary over the A15 phase field.

In the cubic phase range, $B_{c2}(T)$ increases as raising Sn content. There appears an inverse $B_{c2}(T)$ behavior in the tetragonal phase range. A significant influence of the composition concentration on $F_p(B)$ curves is observed. Within the cubic phase range, the increase of Sn content raises the $F_p(B)$ remarkably and shifts the peak in each $F_p(B)$ curve to the right side. The peak shifts to the left and the shape is nearly the same in the tetragonal range. This can be explained by the composition dependencies $B_{c2}(T,\beta)$ and $\kappa(\beta)$.

The effect of composition gradient on the superconducting properties of PIT wires is considered by applying the obtained formulas in Cooley's concentric shells model. The loss of superconductivity near the boundary of A15 layer is mainly associated with the change of the flux-pinning behavior due to the A15 composition variation. The inhomogeneous conductor with gradient severity *N*=20 predicts well the curved "tail" approaching $B_{c2}$ in the Kramer plot. This implies that the composition inhomogeneity is an important factor in the unusual phenomenon of the practical Nb$_3$Sn conductors.

However, we cannot yet include the effect of the alloying addition (Ti and/or Ta) and the matrix material in a ternary Nb$_3$Sn wire. The composition gradient is an important factor in the unusual phenomenon of practical Nb$_3$Sn conductors but not the unique determinant. The most possible role of the alloying addition taken in Nb$_3$Sn is the scattering impurity (Appendix A), which changes the electrical resistivity deeply. We will include this effect in the future work to increase the practical value of the present theory.

**ACKNOWLEDGMENTS**

The authors appreciate the financial supports from the National Natural Science Foundation of China (11372120, 11032006 and 11121202), National Key Project of Magneto-Restriction Fusion Energy Development Program (2013GB110002) and Fundamental Research Funds for the Central Universities (lzujbky-2014-227).

**Appendix A: Scattering by impurity in normal metal**

Let us first introduce the concept of a mean free path $l$, which may be expressed as $l = v\tau$ if $v$ signifies the averaged velocity (approximated as Fermi velocity $v_F$) and $\tau$ the averaged time between collisions. For a gas of free electrons, electrical conduction can be regarded as the diffusion of electrons under an external force $eE$. The mean free path $l$ is a distance travelled by the electrons without undergoing collisions, Fig. A1. The electrical conductivity $\sigma$ is thus associated with $\tau$ in the form of

$$\sigma = e^2 D\upsilon(\mu) = n_e e^2 \tau / m, \tag{A1}$$

where $D$ is the diffusion coefficient $D = lv/3$ and $\upsilon(\mu)$ is the electronic density of states at the Fermi surface [35]. $m$ is the electron mass, $e$ is the unit charge and $n_e$ is the number of electrons per unit volume. The collision depends on certain scattering processes. If the scattering is arisen from impurities and this interaction is elastic and weak, one can obtain the scattering possibility of a system of electrons in the field of impurity centers with Born approximation and the collision integral [36]; consequently, the collision time $\tau$ is obtained as $\tau^{-1} = n_i \lambda$ where

$$\lambda = (4\hbar)^{-1} \upsilon(\mu) \int \left| M_{pp'}(\theta) \right|^2 (1-\cos\theta) d\Omega, \tag{A2}$$

and $n_i$ is the number of impurity atoms per unit volume (i.e. impurity concentration). The integration is taken over the surface of Fermi sphere, $\Omega$ is the solid angle and $\theta$ is the angle between two momentums $\mathbf{p}_0$ ($p_0^2/2m = \mu$ where $p_0$ is the Fermi momentum and $\mu$ is the Fermi energy) and $\mathbf{p}'_0$ on the Fermi surface ($\mathbf{p}$ is fixed and $\mathbf{p}'$ varies with $d\Omega$ during integration). $M_{pp'}(\theta)$ indicates the matrix element of interaction energy of an electron with the impurity.

**Appendix B: Coherence length, penetration depth and GL parameter**

The coherence length $\xi$ indicates a cooper pair coherence between electrons extending to a certain distance in a pure superconductor. $\xi$ is determined as $\xi \sim \hbar v_F / \Delta(T)$ where $\Delta(T)$ is the energy gap in BCS theory [37]. If $T \to T_c$, then $\Delta \approx 3.06[k_B^2 T_c (T_c - T)]^{1/2}$, where $k_B$ is Boltzmann constant (Table A1) [19]. Use is commonly made of the standard coherence length,

$$\xi_0 = \hbar v_F / \pi \Delta(0) = (\eta/\pi^2)\hbar v_F / k_B T_c. \tag{B1}$$

The second equal sign holds since the BCS theory gives [19]

$$k_B T_c = (2\eta/\pi)\hbar \omega_D \exp(-2/g\upsilon) \tag{B2}$$

and

$$\Delta(0) = (\pi/\eta) k_B T_c, \tag{B3}$$

where $\eta = 1.78$. We clarify that, the BCS theory holds true for traditional low-$T_c$ superconductors, with the isotropic model of metal and the weak coupling interaction (i.e. $g\upsilon(\mu) \ll 1$ with the electron-phonon interaction constant $g \sim \hbar^3 / p_0 m$). These conditions are not always fulfilled. For Nb$_3$Sn with $T_c = 4.2$K and $\hbar\omega_D = 94.5$K, one finds $g\upsilon(\mu) = 0.62$ which dose not fulfill $g\upsilon(\mu) \ll 1$. Thus, the BCS theory and its extending results differ from experiment at some extent. However, the disagreement of the theory is remarkably reduced after an appropriate correction [19]. The merit of this theory is the concise physical concepts and brief mathematical representations. The London penetration depth of the magnetic field in a pure superconductor is formulated as $\delta_L = (mc^2/4\pi n_s e^2)^{1/2}$ according to the definition $\delta_L = H_0^{-1} \int_0^\infty H dx$. Here, $n_s$ is known as the number of superconducting electrons per volume in contrast to the total number $n_e$. At the vicinity of $T_c$, there exists $n_s/n_e \approx (T_c - T)/T_c$ upon the Ginzburg-Landau equations. From the above discussion, one figures out that both $\delta_L$ and $\xi$ have a dependence of $(T_c - T)^{-1/2}$ at $T \to T_c$. Hence, the GL parameter $\kappa$, defined as the ratio of $\delta$ and $\xi$, tends to a constant at $T \to T_c$ [38]. For this purpose, we introduce the GL description of $\kappa$,

$$\kappa = 2^{3/2} e H_c \delta^2 / \hbar c, \tag{B4}$$

and the microscopic descriptions of $\delta_L(T)$ [38],

$$\delta_L(T) = \delta_L(0)[1/2(1 - T/T_c)]^{1/2}, \tag{B5}$$

where

$$\delta_L(0) = (mc^2 / 4\pi n_e e^2)^{1/2}. \tag{B6}$$

TABLE A1. Dimension, unit and constant value of physical quantities (Gaussian units).[a]

| Physical quantities | Dimension | Unit | Constant value |
|---|---|---|---|
| Boltzmann constant $k_B$ | $L^2 M T^{-2} T_E^{-1}$ | erg/K | $1.381 \times 10^{-16}$ |
| Critical temperature $T_c$ | $T_E$ | K | - |
| Coefficient of electronic heat capacity $\gamma$ | $L^{-1} M T^{-2} T_E^{-2}$ | erg·cm$^{-3}$·K$^{-2}$ | - |
| Collision time $\tau$ | $T$ | s | - |
| Density of states at Fermi surface $\upsilon(\mu)$ | $L^{-5} M^{-1} T^2$ | erg$^{-1}$·cm$^{-3}$ | - |
| Electron mass $m$ | $M$ | g | $9.109 \times 10^{-28}$ |
| Elementary charge $e$ | $L^{3/2} M^{1/2} T^{-1}$ | esu | $4.803 \times 10^{-10}$ |
| Planck constant $\hbar$ | $L^2 M T^{-1}$ | erg·s | $1.055 \times 10^{-27}$ |
| Electrons number per unit volume $n_e$ | $L^{-3}$ | cm$^{-3}$ | $2.59 \times 10^{23}$ |
| Electrical resistivity $\rho$ | $T$ | s | - |
| Fermi velocity $v_F$ | $L T^{-1}$ | cm/s | - |
| Flux quantum $\Phi_0$ | $L^{3/2} M^{1/2} T^{-1}$ | Maxwell | $\pi \hbar c / e$ |
| Light velocity in vacuum $c$ | $L T^{-1}$ | cm/s | $2.998 \times 10^{10}$ |
| Ratio of Fermi surface $S / S_F$ | Unity | Unity | 0.35 |
| Upper critical field $H_{c2}$ | $L^{-1/2} M^{1/2} T^{-1}$ | Oe | - |
| Upper magnetic induction intensity $B_{c2}$ | $L^{-1/2} M^{1/2} T^{-1}$ | Gauss | - |

[a] All the formula derivations use the Gaussian units except for the unit conversion for practical application. $L$, $M$, $T$ and $T_E$ mean the dimension of length, mass, time and temperature, respectively.

Also the thermodynamic critical field $H_c$ near $T_c$ is formulated as [39]

$$H_c = (16\pi p_0 m / 7 \zeta_3 \hbar^3)^{1/2} k_B T_c (1 - T / T_c), \tag{B7}$$

where the Riemann zeta function $\zeta_3 = \sum_{n=1}^{\infty} n^{-3} \approx 1.2$. Using Eqs. (B5) and (B7) in Eq. (B4) one finds

$$\kappa = (32/7)^{1/2} \pi^{1/2} \zeta_3^{-1/2} \hbar^{-5/2} c^{-1} e k_B m v_F^{1/2} \delta_L^2(0) T_c \approx 3.459 \hbar^{-5/2} c^{-1} e k_B m v_F^{1/2} \delta_L^2(0) T_c. \tag{B8}$$

Substituting $\xi_0$ for $T_c$ by Eq. (B1), we obtain

$$\kappa = (8/7)^{1/2} \eta \pi^{-2} \zeta_3^{-1/2} \hbar^{-3/2} m^{3/2} v_F^{3/2} n_e^{-1/2} \delta_L(0) / \xi_0. \tag{B9}$$

Recall that the Fermi momentum $p_0$ is associated with the number $n_e$ of electrons per volume as [35]

$$p_0 = \hbar(3\pi^2 n_e)^{1/3}, \tag{B10}$$

then we rewrite Eq. (B9) as

$$\kappa = (24/7)^{1/2} \gamma \pi^{-1} \zeta_3^{-1/2} \delta_L(0)/\xi_0 \approx 0.958 \delta_L(0)/\xi_0. \tag{B11}$$

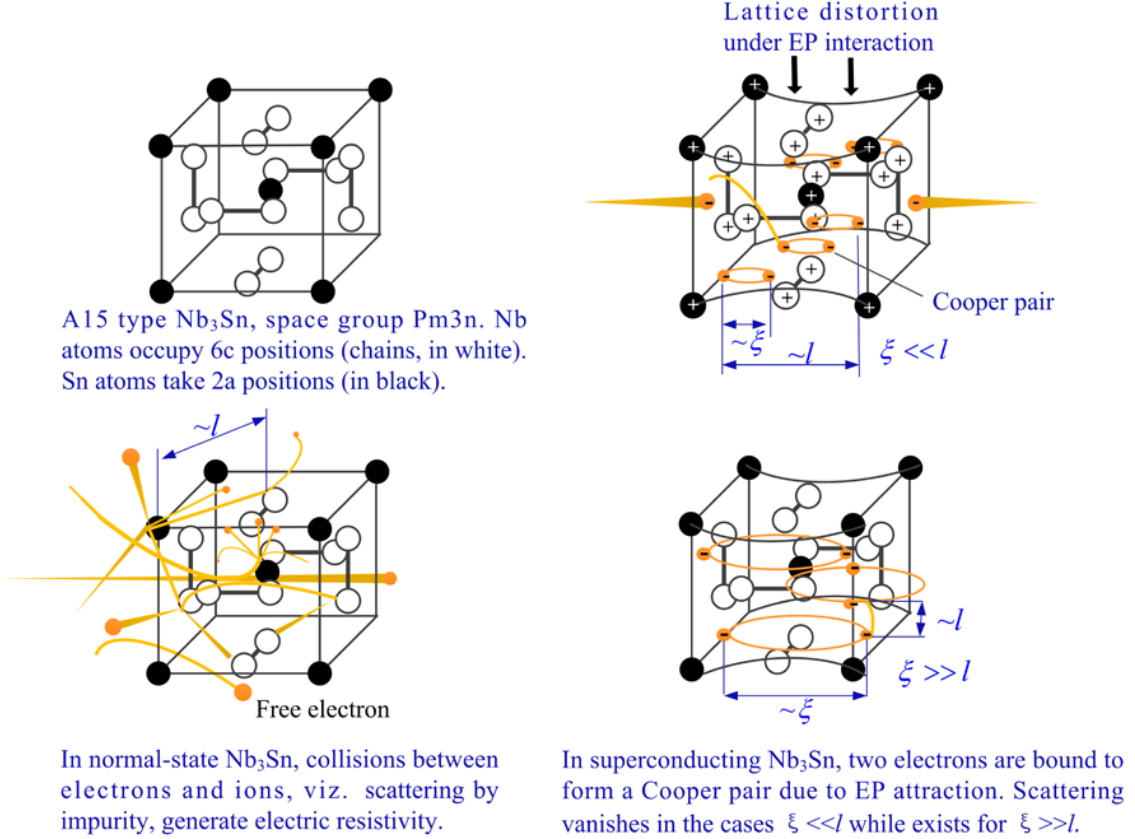

FIG. A1. Schematic of the scattering mechanism in normal-state and superconducting A15-type Nb$_3$Sn.

Extending to a wide temperature range, $\kappa$ is formulated as a temperature-dependent relation $\kappa(T) = \kappa(T_c)\chi_1(T)$; calculations show that $\chi_1(T)$ varies little with $T$ [19]. For the case $\xi \ll l$, scattering from impurity has little impact on the coherence length and the superconductivity [40]. This implies $\kappa$ remains Eq. (B11) for this case.

It is natural to address the opposite limiting case $\xi \gg l$. At first sight, the scattering from impurities reduces the coherence length. Indeed, one could use the diffusion process of an electron in the field of impurity (Appendix A and Fig. A1) to justify the speculation. Recall that under the consideration of scattering, $D$ is the diffusion coefficient with $D = lv/3$ in the diffusion equation $j_e = -D\nabla n_e$. Here, $j_e$ is the diffusion flux, i.e. the number of electrons passing through 1m$^2$ plane in 1s. It follows from the diffusion equation that $x \sim (Dt)^{1/2}$ where $x$ is the distance travelled by a electron in $t$. Without scattering and in the same period $t$, an electron travels a distance of $\xi \sim vt$. Substituting this into $x$ expression, we can obtain the effective coherence length $\xi'$ as $\xi' \sim x \sim (\xi l)^{1/2} \sim (\xi/n_i)^{1/2}$. This is the justification of the above speculation. The scattering of impurity also affects the penetration length $\delta$; the approximation relation is written as $\delta' \sim \delta_L(\xi/l)^{1/2} \sim \delta_L(\xi n_i)^{1/2}$ [19]. Finally, we arrive at the effective GL parameter $\kappa'$ in the impurity field for $\xi \gg l$, $\kappa' = \delta'/\xi' \sim \delta_L/l \sim \delta_L n_i$. Based on a rigorous gauge-invariant solution of the linearized Gor'kov equations [40], one can obtain a precise relation, $\kappa'(T_c) = 0.72\delta_L(0)/l$ and $\kappa'(T) = \kappa'(T_c)\chi_2(T)$. The property of $\chi_2(T)$ is similar with $\chi_1(T)$. We are now ready to summary the dependence of GL parameter $\kappa$ with temperature and impurity concentration $n_i$ [with the aid of Eqs. (B8) and (B11)],

$$\kappa(T) = \kappa(T_c)\chi_1(T), \ \xi \ll l \ (\text{"clean" limit}), \tag{B12}$$

where $\kappa(T_c) \approx 0.96\delta_L(0)/\xi_0$ or $\kappa(T_c) = 3.46\hbar^{-5/2}c^{-1}ek_B m v_F^{1/2}\delta_L^2(0)T_c$.

$$\kappa(T) = \kappa(T_c)\chi_2(T), \ \xi \gg l \ (\text{"dirty" limit}), \tag{B13}$$

where $\kappa(T_c) = 0.72\delta_L(0)\lambda n_i v_F^{-1}$ or $\kappa(T_c) = 0.72\delta_L(0)/l$. We have omitted the superscript of $\kappa'$ in Eqs. (B12) and (B13).

The GL theory of magnetic properties of type II superconductors gives that at the vicinity of $H_{c2}$ [22],

$$H_{c2} = \sqrt{2}\kappa H_c. \tag{B14}$$

**Appendix C: Relation with material parameters**

Now we correlate the electronic density of states, $\upsilon(\mu)$, to the coefficient of electronic heat capacity, $\gamma$, upon isotropic model of metal,

$$\gamma = \pi^2 k_B^2 \upsilon(\mu)/3. \tag{C1}$$

The use of isotropic model equation here is reasonable since the experiments do not show evidence of a highly anisotropic Fermi surface for Nb$_3$Sn [41]. Note that the anisotropy of the physical properties is experimentally identified in compositionally inhomogeneous single crystals [42], scandium [43] and cuprate superconductors [44]. In fact, the electronic structures of Nb$_3$Sn are very complex, and $\upsilon(\mu)$ exhibits a sharp variation relating to the anomalous isotope effect in Nb$_3$Sn superconductors [45]. $\upsilon(\mu)$ may be expressed as a function of the Fermi momentum $p_0$:

$$\upsilon(\mu) = p_0 m/(\pi^2\hbar^3). \tag{C2}$$

Combining Eqs. (C1) and (C2), the Fermi velocity $v_F$ is related to $\gamma$ in a manner of

$$v_F = 3k_B^{-2}m^{-2}\hbar^3\gamma. \tag{C3}$$

Let's consider the electrical resistivity $\rho$, Eq. (A1), and substituting Eqs. (C1) and (C3) into it, one obtains

$$l = 3^{-1}\pi^2 e^{-2} m^2 \hbar^{-3} k_B^4 \gamma^{-2} \rho^{-1}. \tag{C4}$$

This implies a variation of the electronic average free path $l$ as a function of $\gamma$ and $\rho$. We then reformulate the coherence length $\xi_0$, the number of electrons per unit volume, $n_e$, and the London penetration depth $\delta_L(0)$ by using Eqs. (B1), (B6) and (B10) in Eq. (C3), respectively,

$$\xi_0 = 5.34\pi^{-2}m^{-2}k_B^{-3}\hbar^4\gamma T_c^{-1},\tag{C5}$$

$$n_e = 9\pi^{-2}m^{-3}k_B^{-6}\hbar^6\gamma^3,\tag{C6}$$

$$\delta_L(0) = 0.167\pi^{1/2}m^2 ce^{-1}k_B^3\hbar^{-3}\gamma^{-3/2}.\tag{C7}$$

Now using Eqs. (B12), (B13), (C3), (C4) and (C7), we are allowed to find the dependence of $\kappa$ on the three independent material parameters $\gamma$, $\rho$ and $T_c$ for the dirty limit and clean limit near $T_c$:

$$\kappa_{\text{clean}} = 9.43\times 10^{-2}\pi^{3/2}m^4 ce^{-1}k_B^6\hbar^{-7}\chi_1(T)\gamma^{-5/2}T_c,\tag{C8}$$

$$\kappa_{\text{dirty}} = 0.361\pi^{-3/2}cek_B^{-1}\chi_2(T)\gamma^{1/2}\rho.\tag{C9}$$

For a preliminary validation we use the dimensional analysis upon four basic quantities which are length $L$, mass $M$, time $T$ and temperature $T_E$ (Table A1). The LHS of Eq. (C9), for instance, is dimensionless, and the dimension of the RHS is deduced as $[\text{RHS}] = [c][e][k_B]^{-1}[\gamma]^{1/2}[\rho] = (LT^{-1})(L^{3/2}M^{1/2}T^{-1})(L^2MT^{-2}T_E^{-1})^{-1}(L^{-1}MT^{-2}T_E^{-2})^{1/2}T = 1 = [\text{LHS}]$. Further, using Eq. (C3) in Eq. (B7), we express the thermodynamic critical field $H_c$ as

$$H_c = 2.3905\pi^{1/2}T_c(1-T/T_c)\gamma^{1/2}.\tag{C10}$$

Substituting Eqs. (C8)-(C10) into Eq. (B14), one finds the upper critical field $H_{c2}$ in the two limiting cases,

$$H_{c2,\text{clean}} = 0.3188\pi^2 m^4 ce^{-1}k_B^6\hbar^{-7}\chi_1(T)\gamma^{-2}T_c^2(1-T/T_c)\ [\text{Oe}],\tag{C11}$$

$$H_{c2,\text{dirty}} = 1.2204\pi^{-1}cek_B^{-1}\chi_2(T)T_c(1-T/T_c)\gamma\rho\ [\text{Oe}].\tag{C12}$$

For application convenience we convert $\rho$ in Gaussian unit [s] into $\rho_{\Omega m}$ [$\Omega\cdot$m] in fashion to experiments, using the unit conversion relation 1s=1esu$\cdot$cm=$9.0\times 10^{11}\Omega\cdot$cm and $\rho_s = 9^{-1}\times 10^{-9}\rho_{\Omega m}$. For $\gamma$ there is also a conversion relation $\gamma_{\text{erg}} = 1.0\times 10^4\gamma_{\text{mJ}}$ by the relation 1J$\cdot$cm$^{-3}\cdot$K$^{-2}$=$1.0\times 10^7$erg$\cdot$cm$^{-3}\cdot$K$^{-2}$. As experiments always measure $\gamma$ in [mJ$\cdot$mol$^{-1}\cdot$K$^{-2}$], $\gamma_{\text{mJ}}$ [mJ$\cdot$cm$^{-3}\cdot$K$^{-2}$] in the formulas should be further transformed as $\gamma_{\text{mJ}} \to \gamma_{\text{mJ}}/V_{\text{mol}}$ (since $V_{\text{mol}}$mJ$\cdot$cm$^{-3}\cdot$K$^{-2}$ = 1mJ$\cdot$mol$^{-1}\cdot$K$^{-2}$ where $V_{\text{mol}}$ is the volume occupied by 1mol atoms).

For Nb$_3$Sn, recent specific heat measurements reveal $V_{\text{mol}} = 11.085\pm 0.005$ mol$^{-1}$cm$^3$ at 10K temperature [17]. Thus, Eqs. (C11) and (C12) are rewritten as

$$H_{c2,\text{clean}} = C_1\chi_1(T)T_c^2(1-T/T_c)\gamma_{\text{mJ}}^{-2}\ [\text{Oe}],\tag{C13}$$

$$H_{c2,\text{dirty}} = C_2\chi_2(T)T_c(1-T/T_c)\gamma_{\text{mJ}}\rho_{\Omega m}\ [\text{Oe}].\tag{C14}$$

where $C_1 = 0.3188\times 10^{-8}\pi^2 m^4 ce^{-1}k_B^6\hbar^{-7}$ and $C_2 = 1.356\times 10^{-6}\pi^{-1}cek_B^{-1}$. We also concern the determination of whether the experimental sample is in dirty limit or not. The same unit conversations are applied to the electronic free path $l$, the penetration depth $\delta_L(0)$, the coherence length $\xi_0$ and the GL parameter $\kappa$, and we rewrite Eqs. (C4), (C7), (C5), (C8), (C9) and (C10) into

$$l = 30\pi^2 e^{-2}m^2\hbar^{-3}k_B^4\gamma_{\text{mJ}}^{-2}\rho_{\Omega m}^{-1},\tag{C15}$$

$$\xi_0 = 5.34\times 10^4\pi^{-2}m^{-2}k_B^{-3}\hbar^4\gamma_{\text{mJ}}T_c^{-1},\tag{C16}$$

$$\delta_L(0) = 0.167\times 10^{-6}\pi^{1/2}m^2 ce^{-1}k_B^3\hbar^{-3}\gamma_{\text{mJ}}^{-3/2},\tag{C17}$$

$$\kappa_{\text{clean}} = 3.127 \times 10^{-12} \pi^{5/2} m^4 c e^{-1} k_B^6 \hbar^{-7} \chi_1(T) \gamma_{\text{mJ}}^{-5/2} T_c, \tag{C18}$$

$$\kappa_{\text{dirty}} = 4.011 \times 10^{-9} \pi^{-3/2} c e k_B^{-1} \chi_2(T) \gamma_{\text{mJ}}^{1/2} \rho_{\Omega m}, \tag{C19}$$

$$H_c = 2.3905 \times 10^2 \pi^{1/2} T_c (1 - T/T_c) \gamma_{\text{mJ}}^{1/2}. \tag{C20}$$

If experiments give the magnetic flux density $B_{c2}$ [T] then it is necessary to implement $B_{c2} = (H_{c2}[\text{Oe}] \times 10^{-4})$ [T] in the above formulas (since $B_{c2}$ [T]$=\mu_0 \cdot H_{c2}$[A/m] and 1A/m$=4\pi \times 10^{-3}$ Oe with $\mu_0 = 4\pi \times 10^{-7}$ T·m/A).

**Appendix D: Effective electron mass accounting for anisotropic metal**

Accounting for the change of Fermi surface shape due to the transition to the anisotropic model and/or the condensed matter from the isotropic metal model, we employ the formulation of wave vector for the anisotropic metal model:

$$k_F = 3^{1/3} \pi^{2/3} (n_e^{2/3} S/S_F)^{1/2} \tag{D1}$$

with the de Broglie relation $p_0 = m^* v_F = \hbar k_F$ where $m^*$ is the effective electron mass. Here, $S/S_F$ is the ratio of the free Fermi surface to the Fermi surface of a free-electron gas of density $n_e$. Substituting this into Eq. (C3) leads to

$$\begin{aligned} m^* &= 3^{2/3} \pi^{-2/3} k_B^{-2} \hbar^2 (n^{2/3} S/S_F)^{-1/2} \gamma \\ &= 3^{2/3} \times 10^4 \pi^{-2/3} k_B^{-2} \hbar^2 (n^{2/3} S/S_F)^{-1/2} \gamma_{\text{mJ}} \end{aligned}. \tag{D2}$$

Substituting this equation into Eq. (C18) one may obtain

$$\kappa_{\text{clean}} = 5.854 \times 10^5 \pi^{-1/6} c e^{-1} k_B^{-2} \hbar (n^{2/3} S/S_F)^{-2} \chi_1(T) T_c \gamma_{\text{mJ}}^{3/2}. \tag{D3}$$

The following equations can be obtained in the same way:

$$l = 9 \times 10^9 (3\pi^2)^{1/3} \hbar e^{-2} (n^{2/3} S/S_F)^{-1} \rho_{\Omega m}^{-1}, \tag{D4}$$

$$\xi_0 = 1.234 \times 10^{-4} \pi^{-2/3} k_B (n^{2/3} S/S_F) \gamma_{\text{mJ}}^{-1} T_c^{-1}, \tag{D5}$$

$$\delta_L(0) = 72.257 \pi^{-5/6} c e^{-1} \hbar k_B^{-1} (n^{2/3} S/S_F)^{-1} \gamma_{\text{mJ}}^{1/2}, \tag{D6}$$

$$H_{c2,\text{clean}} = 1.979 \times 10^8 \pi^{1/3} c e^{-1} k_B^{-2} \hbar (n^{2/3} S/S_F)^{-2} \chi_1(T) \gamma_{\text{mJ}}^2 T_c^2 (1 - T/T_c). \tag{D7}$$

**Appendix E: Lambert $W(X)$ function**

Equation (4) can be solved via the Lambert $W(X)$ function, which is defined as the solution to $W \exp(W) = X$. $W(K, X)$ is the $K$-th branch of the multi-valued function $W(X)$. It follows that a solution to $y = x^{-2} \ln x$ is $x = \exp[-W(K, -2y)/2]$, where $K$ representing the branch of the multivalued $W$ is selected to ensure positive and real $x$. For $\omega_0 k_B^{-1} T_c^{-1}$ in question, we arrive at $\omega_0 k_B^{-1} T_c^{-1} = \exp\{-0.5W[K, -0.377(\eta_{\Delta(0)} - 1)]\}$ with $K = -1$.